\documentclass[twocolumn,showpacs,prl,aps,tightenlines,superscriptaddress]{revtex4}

\usepackage{graphics,graphicx}
\usepackage{amsmath}
\usepackage{psfrag}
\usepackage{dcolumn}
\usepackage{psfrag}

\newcommand{\be}{\begin{equation}}
\newcommand{\ee}{\end{equation}}
\newcommand{\bea}{\begin{eqnarray}}
\newcommand{\eea}{\end{eqnarray}}

\begin{document}

\title{
New Class of Gravitational Wave Templates for 
Inspiralling 
Compact Binaries 
}

\author{ Achamveedu Gopakumar
}
\email{A.Gopakumar@uni-jena.de}
\affiliation{Theoretisch-Physikalisches Institut,
Friedrich-Schiller-Universit\"at Jena,
Max-Wien-Platz 1,
07743 Jena, Germany}

\date{\today}

\begin{abstract}
Compact binaries inspiralling along  quasi-circular orbits are the
most plausible gravitational wave (GW)
sources  for the operational, planned and proposed laser interferometers. 
We provide new class of restricted post-Newtonian accurate GW templates for 
non-spinning compact binaries inspiralling along PN accurate quasi-circular orbits.
Arguments based on data analysis, theoretical and astrophysical considerations are
invoked to show why these time-domain Taylor approximants should be interesting to various 
GW data analysis communities. \footnote{This is roughly the version submiited to 
Phys.\ Rev.\ Lett. on 18/08/2007. A substantially
modified version that also addresses objections
of the referee is under preparation.}

\end{abstract}


\maketitle

\paragraph{Introduction.---}

Inspiralling comparable mass compact binaries are the most plausible sources of
gravitational radiation for the 
operational, planned and proposed 
laser interferometric
GW interferometers.
GW data analysts,
analyzing noisy data from the interferometers,  require accurate and
efficient temporally evolving GW
polarizations, $h_{+} (t)$ and $h_{\times}(t)$,
the so-called GW search templates.  
It is expected that weak GW signals, buried in the noisy interferometric data,
should be extracted by employing the 
technique of `matched filtering'. This is an optimal technique if and only if 
one can construct search templates that
accurately model expected GW signals from astrophysical sources,
especially in their phase evolution.
Till the late stages of binary inspiral, GW signals may be accurately modeled 
using the post-Newtonian (PN)
approximation to general relativity.
The PN approximation to the dynamics of inspiralling compact binaries,
usually modeled to consist of point masses,
provides, for example,
the equations of motion  as corrections to the Newtonian one
in terms of
 $({v}/{c})^2 \sim {G m}/{c^2\,r}$,
where $v$, $m$, and $r$ are
the characteristic orbital velocity,
the total mass, and the typical orbital separation,
respectively. In PN computations, it is customary to treat a
non-spinning inspiralling compact binary to consist of two point masses moving in quasi-circular 
orbits. These PN computations, to date provided {\em four} quantities that are required to 
do astrophysics with GW interferometers. For inspiralling compact binaries, the relevant four
quantities are the 3PN accurate dynamical (orbital) energy ${\cal E}(x)$,
expressed as a PN series in terms 
of $x = \left ( G\, m\, \omega_{\rm 3PN} /c^3 \right )^{2/3}$, $ \omega_{\rm 3PN}(t) $ being the 
3PN accurate orbital angular frequency, the 3.5PN accurate expression for GW energy luminosity
$ {\cal L }(x) $ and the 2.5PN amplitude corrected  expressions for $h_{+} (t)$ and $h_{\times}(t)$,
written in terms of the orbital phase $\phi$ and $x$ \cite{BDI}.

   GW data analysts employ these inputs to construct various types of search templates and let us
take a closer look at the so-called TaylorT1 and TaylorT2 waveforms implemented
in the LSC Algorithms Library (LAL) \cite{LAL}.
These two template families employ the following expression for the so-called restricted PN waveform
\be
h(t) \propto \left ( \frac{G\,m \, \omega (t) }{c^3} \right )^{2/3} \, \cos 2\, \phi(t)\,,
\label{Eq.I1}
\ee
where the proportionality constant 
may be set to unity for non-spinning compact binaries. 
At a given PN order, the above mentioned two families provide two slightly different ways to   
compute $\omega (t)$ and $\phi(t)$.
The TaylorT1 family numerically solves the following two differential equations:
\be
\frac{d \phi (t)}{dt} = \omega (t)\,;  \,\,\, 
\frac{d\,\omega (t)}{dt} =  -{\cal L}( \omega) \bigg / \frac{ d {\cal E}}{d \omega}\,,
\label{EqI2}
\ee
where, for example, ${\cal L}( \omega)$ and ${\cal E}$ are 
respectively the
3.5PN accurate 
GW energy luminosity and the 3PN accurate orbital energy for TaylorT1 3.5PN waveforms.
In other words, for a given PN member of the TaylorT1  family, $ \omega(t) $ and $\phi(t)$ 
are  computed by numerically solving the related approximants in Eq.~(\ref{EqI2}).
To construct a member of TaylorT2 family, say TaylorT2 3.5PN, we require 3.5PN (Taylor expanded)
accurate version of $d\, \omega (t)/dt $, appearing in Eq.~(\ref{EqI2}).
The differential equations that define  $\omega(t) $ and $\phi(t)$
for TaylorT2 3.5PN  waveforms can be symbolically displayed as
\begin{align}
\label{EqI3}
\frac{d \phi (t)}{dt} &= \omega (t)\,;  
\frac{d\,\omega (t)}{dt} =  \frac{96}{5}
\left ( \frac{ G\, {\cal M}\,\omega}{c^3} \right )^{5/3}
\omega^2 \, \biggl \{ 1 
\nonumber
\\
& \quad
+ {\cal O}(\nu) 
+ {\cal O}(\nu^{3/2}) 
+ {\cal O}(\nu^{2})
+ {\cal O}(\nu^{5/2}) 
\nonumber
\\
& \quad
+ {\cal O}(\nu^{3}) 
+ {\cal O}(\nu^{7/2}) \biggr \}\,,
\end{align}
where $ \nu = 1/c^2$ is a PN ordering parameter and the explicit expressions for these PN 
contributions may be extracted
from Refs.~\cite{BDI}. In the above equation, the chirp mass ${\cal M} \equiv m\,\eta^{3/5}$, where 
$\eta$ is the usual symmetric mass ratio and $m$ being the total mass of the binary.

   In this paper, we provide prescriptions to compute {\em three} new types of 
time-domain Taylor approximants that should be, in our opinion,
interesting to various  GW data analysis 
communities. Let us first list the salient features of these new templates that also employ 
an expression similar to
Eq.~(\ref{Eq.I1}) to generate waveforms. The {\em three} important features of our Taylor approximants 
are the following.
The first point is that, in comparison with TaylorT1 and Taylor T2 waveforms,
for a given GW frequency window and at a given PN order, our prescriptions 
will provide more accumulated GW cycles.
Further, our approaches to compute $h(t)$ are numerically as cheap (expensive) as 
TaylorT1 and Taylor T2 waveforms.
Let us consider the second point. It is desirable to
construct GW templates using
the mathematical formulation employed to construct
the (heavily employed) PN
accurate relativistic
\emph{Damour-Deruelle timing formula}
for binary pulsars \cite{DD86}. This is because formally GW phasing for inspiralling compact binaries
and timing of relativistic binary pulsars are quite similar.
Our construction of these new Taylor approximants are indeed influenced by the GW phasing
formalism, available in Ref.~\cite{DGI}, that provided a method to construct
GW templates for compact binaries of arbitrary mass ratio
moving in inspiralling eccentric orbits.
We recall that
the techniques adapted in Ref.~\cite{DGI}
were influenced by the mathematical formulation,
developed in Ref.~\cite{TD82}, to compute the 
\emph{Damour-Deruelle timing formula}.
Finally, 
a recent preliminary investigation indicates that our new  Taylor approximants, at the dominant
radiation reaction order,  should be 
very efficient  in capturing GWs from compact binaries inspiralling along PN accurate and mildly 
eccentric orbits \cite{TG07}. This is, in our opinion, a very attractive feature
for GW data analysts
as GWs from inspiralling (astrophysical) compact binaries should have some tiny 
eccentricities, when their GWs enter the bandwidth of laser interferometers.

 Let us describe how we construct these new types of PN accurate time-domain
Taylor approximant GW search templates.
GWs from inspiralling (astrophysical) compact binaries will have some tiny
eccentricities around orbital frequencies of $20$Hz. For example, using 
Ref.~\cite{DGI}, it is not that difficult to show that 
the orbital eccentricity of the Hulse-Taylor binary
pulsar when its orbital frequency reaches around $20$ Hz will be $\sim 10^{-6}$.
Therefore, let us take a closer look at how one can describe, 
in a PN accurate manner, eccentric orbits, motivated by the fact that 
GW phasing requires accurate orbital description.
Couple of decades ago,
it was demonstrated that  associated with a PN accurate non-circular orbit, there
exists {\em two gauge invariant} quantities, if expressed in terms the conserved orbital 
energy and angular momentum of the binary \cite{DS88}.
These are the PN accurate mean motion $n$ and $k$ that measures the advance of periastron 
in the time interval $T$,
$T$ being the radial orbital period such that $n =2\,\pi/ T $. It is quite convenient to define 
these quantities in the PN accurate Keplerian type parametric solution to the 
conservative PN accurate compact
binary dynamics, available in Refs.~\cite{DD}.
When eccentricity parameter, say time eccentricity $e_t$, associated with the PN accurate 
Keplerian type parametrization approaches zero, one can define PN accurate orbital 
angular frequency  $ \omega \equiv d \phi/dt = n \left( 1 + k \right )$.
This implies that for PN accurate circular orbits, the angular part of the orbital motion is
simply given by $ \phi - \phi_{0} = n \times (1 + k) \times ( t - t_0) $.
To 3PN order, using Ref.~\cite{KG06}, in the limit $ e_t \rightarrow 0$, we have
\begin{align}
\omega_{\rm 3PN} = & n \biggl \{ 
1  
+3\,{\xi}^{2/3}
+ \left( {
\frac {39}{2}}-7\,\eta \right) {\xi}^{4/3}
+ \biggl [ {\frac {315}{2}}
\nonumber
\\
& \quad
+7\,{\eta}^{2}
+ \biggl ( -{\frac {817}{4}}
+{
\frac {123}{32}}\,{\pi }^{2} \biggr ) \eta \biggr ] {\xi}^{2}
\biggr \}\,,
\label{Eq2.1}
\end{align}
where $ \xi = G\,m\, n/c^3$.
Let us now compute employing 
PN accurate expressions for 
${\cal E}(x)$ and $ {\cal L}(x)$, available in Refs.~\cite{BDI}, the following 3PN 
accurate expression for the orbital energy ${\cal E}$ and 3.5PN accurate 
GW energy luminosity ${\cal L}$, in terms of $ \xi$, as
\begin{subequations}
\label{Eq2.2}
\begin{align}
{\cal \tilde E}(\xi) &= \xi^{2/3} \biggl \{
1
+ \left[ \frac{5}{4} - \frac{ \eta}{12} \right] {\xi}^{2/3}
+ \biggl [{\frac {45}{8}}-{ \frac {21}{8}}\,\eta
\nonumber
\\
& \quad
-\frac{1}{24}\,{\eta}^{2}  
\biggr ] {\xi}^{4/3}
+ 
\biggl [ 
{\frac {7975}{192}}-{\frac {35}{5184}}\,{\eta}^{3}+{\frac {
1031}{288}}\,{\eta}^{2}
\nonumber
\\
& \quad
+ \left( -{\frac {30403}{576}}+{\frac {41}{96}}
\,{\pi }^{2} \right) \eta 
\biggr ]
 {\xi}^{2}
\biggr \}\,,
\\
{\cal L}(\xi) &= \frac{32}{5} \, \eta^2 \xi^{10/3} \,
\biggl \{
1+ 
 \left( -{\frac {35}{12}}\, \eta+{\frac {2113}{336}} \right) {\xi}^{2/3}
\nonumber
\\
& \quad
+4\,\pi\,\xi
+ \left( 
{\frac {458461}{9072}}-{\frac {20129}{504}}\,\eta+{\frac {65}{18}}\,{
\eta}^{2} \right) {\xi}^{4/3}
\nonumber
\\
& \quad
+ \left( -{\frac {583}{24}}\,\eta+{\frac {26753}{672}} \right) \pi\,{\xi}^{5/3}
+ \biggl [ 
\biggl ( \frac{16}{3}
\nonumber
\\
& \quad
+{\frac {41}{3}}\,\eta \biggr ) {\pi}^{2}
+{\frac {13106635373}{23284800}} 
-{\frac {6881951}{7776}}\, \eta
\nonumber
\\
& \quad
+{ \frac {375997}{3024}}\,{\eta}^{2} 
-{\frac {775}{324}}\,{\eta}^{3}
-{\frac {1712}{105}}\,\biggl ( \gamma 
+ \log(4\, \xi^{1/3}) \biggr )
\biggr ] {\xi}^{2}
\nonumber
\\
& \quad
+ \biggl ( {\frac {771833}{2016}}-{\frac {624559}{1728}}\,\eta
+{\frac { 193385}{3024}}\,{\eta}^{2} \biggr )  \pi\,{\xi}^{7/3}
\biggr \}\,,
\end{align}
\end{subequations}
where ${\cal \tilde E}=-2\,E$, $E$ being the dimensionless non-relativistic energy 
per unit reduced mass \cite{DGI} and $\gamma$ being the Euler's gamma.
We are now in a position to construct, 
in our terminology,
TaylorK1 3.5PN and 
TaylorK2 3.5PN  restricted PN waveforms. In our approach, the form of the 
restricted PN waveform, Eq.~(\ref{Eq.I1}), becomes
$ h(t) \propto \left ( \frac{G\,m \, n(t) }{c^3} \right )^{2/3} \, \cos 2\, \phi(t)\,,$
This is allowed because at Newtonian order
$\omega = n$ 
and the amplitude is indeed Newtonian accurate in
Eq.~(\ref{Eq.I1}). For TaylorK1 3.5PN accurate waveform, $n(t)$ and $\phi(t)$ are numerically
obtained using the following two differential equations
\begin{subequations}
\label{Eq2.4}
\begin{align}
\frac{d \phi}{dt} &= \omega_{\rm 3PN}\,,
\label{Eq2.4a}
\\
\frac{d n}{dt} &= - {\cal L}(\xi) \bigg / \frac{d { E}}{d n}
\label{Eq2.4b}
\end{align}
\end{subequations}
To construct our TaylorK2 3.5PN waveforms, as expected, we Taylor expand, in terms of $\xi$,
the RHS of Eq.~(\ref{Eq2.4b}) and this leads to
\begin{subequations}
\label{Eq2.6}
\begin{align}
\frac{d \phi}{dt} &= \omega_{\rm 3PN}\,,
\label{Eq2.6a}
\\
\frac{d n}{dt} &=  {\frac {96}{5}}\,\eta\,{n}^{2}{\xi}^{5/3} 
\biggl \{
1
+ \left( {\frac {1273}{336 }}-\frac{11}{4}\,\eta \right) {\xi}^{2/3}
+4\,\pi\,\xi
\nonumber
\\
& \quad
+ \left( {\frac {438887}{18144}}+{\frac {59}{18}}\,{\eta}^{2}
-{\frac {49507}{ 2016}}\,\eta \right) {\xi}^{4/3}
+ \biggl ( {\frac {20033}{672}}
\nonumber
\\
& \quad
-{\frac {189}{8}}\,\eta \biggr ) \pi\,{\xi}^{5/3}
+ \biggl [ 
 \left( 
\frac{16}{3} 
+{\frac {287}{24}}\,\eta \right) {\pi}^{2}
\nonumber
\\
& \quad
-{\frac {5605}{2592}}\,{\eta}^{3}
+{\frac {617285}{8064}} \,{\eta}^{2}
-{\frac {16554367}{31104} }\,\eta
\nonumber
\\
& \quad
+{\frac {38047038863}{139708800}}
-{\frac {1712}{105}}\,\left ( \gamma + \log 4\,\xi^{1/3} \right )
\biggr ] \xi^2
\nonumber
\\
& \quad
+ \biggl [
 {\frac {91495}{1512}}\,{\eta}^{2}
-{\frac {1608185}{6048}}\,\eta 
+{\frac {971011}{4032}}
\biggr ] \pi\,{\xi}^{7/3}
%
%
\biggr \}\,,
\label{Eq2.6b}
\end{align}
\end{subequations}
Let us now specify, for example,
 the limits of integration for $n$ to construct TaylorK1 3.5PN  and TaylorK2 3.5PN
waveforms. For initial LIGO, it is customary to use $\omega_i$ and $\omega_f$, the initial and final 
final values of $\omega$, to be $40\, \pi$ Hz and
$(6^{(3/2)} \, m )^{-1}$Hz, where $\omega_f$ 
is twice the conventional orbital angular frequency of the innermost stable circular orbit
for a test particle around a  Schwarzschild black hole. 
With these inputs, the initial and final values of $n$, denoted by $n_i$ and $n_f$,  are 
numerically computed using Eq.~(\ref{Eq2.1}). 
This is justified because of the observation in Ref.~\cite{TG06} that the 
quadrupolar GW frequency from a compact binary,
having PN accurate orbital motion, appears at
$( 1 + k) \, n/ \pi$.
At 3PN order, for a compact binary having 
$m= 11.4 M_{\odot} $ and $\eta \sim 0.108$
we have $n_i \sim 
111.32$ Hz and $n_f \sim 679.3$Hz.
In our approaches to construct, for example, TaylorK1 2PN  and TaylorK2 2PN
waveforms, we use only the 2PN accurate relation connecting $\omega$ and $n$.
%
  
    Let us now compute in the time domain 
the accumulated number of GW cycles,
$\mathcal N_{GW}$,
in a given GW frequency window, 
by numerically integrating    
Eqs.~(\ref{Eq2.4}) and (\ref{Eq2.6}) representing temporal evolutions for 
TaylorK1 and TaylorK2 waveforms at
{\em four} different PN orders, namely 
 2PN, 2.5PN, 3PN and 3.5PN orders,  for three canonical compact
binaries usually considered in the GW literature [we restrict these orbital evolutions 
such that emitted GWs are in the GW frequency window defined by $40$ Hz and 
$ ( 6^{3/2} \, \pi\, m )^{-1}$Hz].
Let us also compare these $\mathcal N_{GW}$ with what is expected from 
TaylorT1 and TaylorT2 waveforms at these four different PN orders . The numbers, relevant for 
initial LIGO,  are listed in
Table~\ref{tab1} where we compare $\mathcal N_{GW}$ resulting from
TaylorK2 and TaylorT2 prescriptions [results are similar while comparing TaylorK1 and TaylorT1]
\begin{table}[!ht]
\caption{
\label{tab1}
Accumulated number of GW cycles, relevant for initial LIGO, 
for three types of canonical binaries at four different PN orders using 
TaylorK2 and TaylorT2 waveforms.
The values of $\mathcal N_{GW}$ arising from TaylorT2 waveforms are given in parentheses.
We note that TaylorK2 waveforms provide more $\mathcal N_{GW}$ compared to 
TaylorT2 waveforms. For high mass binaries, the convergence of $\mathcal N_{GW}$ is not that
pronounced for TaylorK2 waveforms compared to TaylorT2 waveforms.
}
\begin{tabular}{||l|r|r|r|}
\hline
 $m_1/ M_{\odot} : m_2/ M_{\odot} $ & $1.4 : 1.4$ & $1.4 : 10$ & $10 : 10$ \\
\hline
$ {\rm 2PN} $ \hfill          &  1616.4 (1613.5)  & 345.6 (333.8)  &  57 (52.6) \\
${\rm 2.5PN } $ \hfill        & 1613.5 (1605.8) & 333.8 (333.1) & 53.8 ( 52.6)  \\
${\rm 3PN}$ \hfill          & 1623.4 (1616) & 347.3 ( 330.9) & 57.6 (52.9)  \\
${\rm 3.5PN} $ \hfill          & 1620.6 (1615.4) & 342.4 (330.5) & 56.2 (52.5) \\
\hline
\end{tabular}
\end{table}

   We are aware that LAL also provides routines to create TaylorT3 waveforms. In this prescription,
both $\phi(t)$ and $\omega(t)$, appearing in Eq.~(\ref{Eq.I1}), are 
given as explicit  PN accurate functions of time.
 These explicit time dependencies are usually expressed in terms of the so-called
`adimensional' time variable $ \theta = \frac{c^3\, \eta}{5\, G\, m} \left( t_c - t \right)$,
where $t_c$ is the PN accurate coalescence time. 
It is indeed  possible for us to
compute $n(t)$, using Eqs.~(\ref{Eq2.6}),  as a PN series in terms of $\theta$.
However, we are reluctant to
repeat what is done in TaylorT3 waveforms to get $\phi(t)$ with the help of Eqs.~(\ref{Eq2.6}).
Observe that  radiation reaction and hence temporal evolution of $n$ first appears at 
2.5PN order and therefore, in our opinion,
it is better to keep $d \phi/dt$ to at least
2PN order in Eqs.~(\ref{Eq2.6}) to be consistent in a PN way[see Refs.~\cite{DGI,TG07} where similar 
approaches are employed].

  It is important to note, while constructing these time-domain Taylor waveforms, that we employed
the following two arguments.
The first one is the standard argument that equates the rate
of decrease of the conserved orbital energy of a compact binary to the opposite of GW
luminosity.  
However, for constructing TaylorT1, TaylorT2, TaylorK1 and TaylorK2
waveforms, one requires
additional PN accurate relations relating $\omega$ (or $n$ as the case may be)
to the conserved orbital energy.
Further, we speculate
that the two different ways of computing $d \omega/dt$, enforced in TaylorT1 and TaylorT2
waveforms, may be based on the fact that
observationally $d \omega/dt$
(or the above mentioned standard argument) is only tested to the
Newtonian radiation reaction order by the accurate timing of binary pulsars.
Therefore, it is natural to ask if we can construct $h(t)$ employing 
only the energy balance argument.
This is indeed possible as demonstrated below
\begin{subequations}
\label{Eq2.7}
\begin{align}
h(\hat t) & \propto {\cal \tilde E}(\hat t) \, \cos 2\,\phi (\hat t) \,,
\label{Eq2.7a}
\\
\frac{d \phi}{d \hat t}  &= \zeta^{3/2} \biggl \{
1+ 
 \frac{1}{8} \left[ {9}+\eta \right] \zeta
+ \biggl [
{\frac {891}{128}}
-{\frac {201 }{64}}\,\eta
+{\frac {11}{128}}\,{\eta}^{2} 
\biggr ] {\zeta}^{2}
\nonumber
\\
& \quad
+
\biggl [
{\frac {41445}{1024}}
+ \left( -{\frac {309715}{3072}}
+{\frac {205}
{64}}\,{\pi}^{2} \right) \eta 
+{\frac {1215}{1024}}\,{\eta}^{2}
\nonumber
\\
& \quad
+{\frac {45}{1024}}\,{\eta}^{3}
\biggr ] {\zeta}^{3}
%
\biggr \}\,,
\label{Eq2.7b}
\\
\frac{d \zeta}{d \hat t} &=
{\frac {64}{5}}\,\eta\,{\zeta}^{5} 
\biggl \{
1+ \left[ {\frac {13}{336}}-\frac{5}{2}\,\eta \right] \zeta
+4\,\pi\,{ \zeta}^{3/2}
\nonumber
\\
& \quad
+ \left[ 
{ \frac {117857}{18144}}
-{\frac {12017}{2016}}\,\eta
+\frac{5}{2}\,{\eta}^{2} 
\right] {\zeta}^{2}
+ \biggl [
{\frac {4913}{672 }} 
\nonumber
\\
& \quad
-{\frac {177}{8}}\,\eta
\biggr ] \pi\,{\zeta}^{5/2}
+ \biggl [ 
\left( {\frac {369}{32}}\,\eta+ \frac{16}{3}\right) {\pi}^{2}
\nonumber
\\
& \quad
+{\frac {37999588601}{279417600}}
-{\frac {24861497}{72576}}\,\eta
+{\frac {488849}{16128}}\,{\eta}^{2}
\nonumber
\\
& \quad
-{\frac {85}{64} }\,{\eta}^{3}
-{\frac {1712}{105}}\,
\biggl ( \ln  \left( 4\,\sqrt {\zeta} \right) + \gamma \biggr )
 \biggr ] {\zeta}^{3}
\nonumber
\\
& \quad
+ \biggl [
 {\frac {613373}{12096}}\,{\eta}^{2}+{\frac {129817}{2304}}-{
\frac {3207739}{48384}}\,\eta 
\biggr ] \pi\,{\zeta}^{7/2}
\biggr \}\,,
%
%
%
%
%
%
\label{Eq2.7c}
\end{align}
\end{subequations}
where $ \hat t = t\, c^3/G\,m $ and $\zeta = {\cal \tilde E}$.
We call the resulting $h(\hat t)$  as TaylorEt waveforms. 
The values of $\zeta $ corresponding to  $\omega_i$ 
and $\omega_f$ can numerically evaluated using the RHS of Eq.~(\ref{Eq2.7b}) for 
$d \phi/ d \hat t = \hat \omega$.

  We  evaluated ${\cal N}_{GW}$ associated with TaylorEt 3.5PN waveforms for the three 
canonical compact binaries  and the numbers are the following.
For neutron star binaries, $m = 2.8 M_{\odot}$ and $ \eta = 0.25$, $ {\cal N}_{GW} =1617.4$ and for 
the usual black hole-neutron star binaries, $ m = 11.4  M_{\odot}$ and $\eta =0.108$,
we have $ {\cal N}_{GW} =335.4$. For typical stellar mass black hole binaries,
$ m = 20  M_{\odot}$ and $\eta =0.25$, one gets $ {\cal N}_{GW} =54.0$.
It is interesting to note that we get larger ${\cal N}_{GW}$, compared to TaylorT 3.5PN 
waveforms and lower values compared to TaylorK 3.5PN waveforms.
It should be related to the fact that it takes more time for TaylorEt prescription to 
reach $\omega_f$ form $\omega_i$ compared to TaylorT1 (or TaylorT2) approach and the 
opposite is true for the cases of 
TaylorK1 (or TaylorK2).  The observation that TaylorEt waveforms also provide
more number of GW cycles in a give GW frequency window, in our opinion, makes 
it our third prescription to compute $h(t)$.

\paragraph{Conclusions.---}

We provided new ways of constructing restricted time-domain PN accurate waveforms for
non-spinning compact binaries 
inspiralling along PN accurate quasi-circular orbits. Our prescriptions employed
PN accurate expressions for the conserved orbital energy and GW luminosity, 
available in Refs.~\cite{BDI}, in a democratic manner and heavily 
depended on certain PN accurate gauge invariant quantities, first introduced in Ref.~\cite{DS88}.
These template waveforms provide more number of 
accumulated
GW cycles in a given GW frequency window 
and may be useful in detecting GWs from inspiraling compact binaries that should have 
`teeny-weeny' orbital eccentricities. Further, our approaches are influenced by the 
way PN accurate \emph{Damour-Deruelle timing formula}
was constructed.  Therefore, we feel that our TaylorK1, TaylorK2 and TaylorEt waveforms
should be of certain interest to the practitioners of LAL.
Further, we feel that our restricted PN waveforms should be useful for the
the recently initiated {\it mock LISA data challenge} task force.

 The data analysis implications of these templates,
relevant for both ground and space 
based GW detectors, are under active investigations in collaborations 
with Stas Babak, Sukanta Bose, Christian R\"over and Manuel Tessmer.
The GW phase evolution under our prescription  is also being
compared with its counterpart in
numerical relativity based binary black inspiral.

\acknowledgments

I am indebted to Gerhard Sch\"afer for
illuminating discussions and persistent encouragements.
Lively discussions with Manuel Tessmer are warmly acknowledged.
This work is supported in part by
the DFG (Deutsche Forschungsgemeinschaft) through SFB/TR7
``Gravitationswellenastronomie'' and
the DLR (Deutsches Zentrum f\"ur Luft- und Raumfahrt) through ``LISA Germany''.


\end{document}